\documentclass[sigconf]{acmart}

\AtBeginDocument{%
  \providecommand\BibTeX{{%
    \normalfont B\kern-0.5em{\scshape i\kern-0.25em b}\kern-0.8em\TeX}}}

\setcopyright{acmlicensed}
\copyrightyear{2018}
\acmYear{2018}
\acmDOI{XXXXXXX.XXXXXXX}

\acmConference[Conference acronym 'XX]{Make sure to enter the correct
  conference title from your rights confirmation emai}{June 03--05,
  2018}{Woodstock, NY}
\acmBooktitle{Woodstock '18: ACM Symposium on Neural Gaze Detection,
 June 03--05, 2018, Woodstock, NY} 
\acmISBN{978-1-4503-XXXX-X/18/06}

\usepackage{xspace}
\usepackage{soul}
\usepackage{comment}

\usepackage{booktabs}
\usepackage{multirow}

\newcommand{\eg}{{\it e.g.}\xspace}
\newcommand{\ie}{{\it i.e.}\xspace}

\newcommand{\system}{\textsc{SciCapenter}\xspace}

\usepackage{cleveref}

\copyrightyear{2024} 
\acmYear{2024} 
\setcopyright{rightsretained} 
\acmConference[CHI EA '24]{Extended Abstracts of the CHI Conference on Human Factors in Computing Systems}{May 11--16, 2024}{Honolulu, HI, USA}
\acmBooktitle{Extended Abstracts of the CHI Conference on Human Factors in Computing Systems (CHI EA '24), May 11--16, 2024, Honolulu, HI, USA}
\acmDOI{10.1145/3613905.3650738}
\acmISBN{979-8-4007-0331-7/24/05}

\begin{document}

\title[\system]{\system: Supporting Caption Composition for Scientific Figures with Machine-Generated Captions and Ratings}

% \title{The Name of the Title is Hope}

\author{Ting-Yao Hsu}
\orcid{1234-5678-9012}
\affiliation{%
  \institution{The Pennsylvania State University}
  % \streetaddress{P.O. Box 1212}
  \city{University Park}
  \state{PA}
  \country{USA}
}
\email{txh357@psu.edu}

\author{Chieh-Yang Huang}
\affiliation{%
  \institution{The Pennsylvania State University}
  % \streetaddress{P.O. Box 1212}
  \city{University Park}
  \state{PA}
  \country{USA}
}
\email{chiehyang@alumni.psu.edu}

\author{Shih-Hong Huang}
\affiliation{%
    \institution{The Pennsylvania State University}
  % \streetaddress{P.O. Box 1212}
  \city{University Park}
  \state{PA}
  \country{USA}
}
\email{szh277@psu.edu}

\author{Ryan Rossi}
\affiliation{%
 \institution{Adobe Research}
 \city{San Jose}
 \state{CA}
 \country{USA}}
\email{ryrossi@adobe.com}

\author{Sungchul Kim}
\affiliation{%
  \institution{Adobe Research}
 \city{San Jose}
 \state{CA}
 \country{USA}}
\email{sukim@adobe.com}

\author{Tong Yu}
\affiliation{%
  \institution{Adobe Research}
 \city{San Jose}
 \state{CA}
 \country{USA}}
\email{tyu@adobe.com}

\author{Clyde Lee Giles}
\affiliation{%
  \institution{The Pennsylvania State University}
  % \streetaddress{P.O. Box 1212}
  \city{University Park}
  \state{PA}
  \country{USA}
}
\email{clg20@psu.edu}

\author{Ting-Hao `Kenneth' Huang}
\affiliation{%
  \institution{The Pennsylvania State University}
  % \streetaddress{P.O. Box 1212}
  \city{University Park}
  \state{PA}
  \country{USA}
}
\email{txh710@psu.edu}

\renewcommand{\shortauthors}{Trovato and Tobin, et al.}

\begin{abstract}
  
Crafting effective captions for figures %in scholarly documents 
is important.
Readers heavily depend on these captions to grasp the figure's message.
However, despite a well-developed set of AI technologies for figures and captions, these have rarely been tested for usefulness in aiding caption writing.
This paper introduces \system, an interactive system that puts together cutting-edge AI technologies for scientific figure captions to aid caption composition.
\system generates a variety of captions for each figure in a scholarly article, providing scores and a comprehensive checklist to assess caption quality across multiple critical aspects, such as helpfulness, OCR mention, key takeaways, and visual properties reference. 
Users can directly edit captions in \system, resubmit for revised evaluations, and iteratively refine them. 
A user study with Ph.D. students indicates that \system significantly lowers the cognitive load of caption writing.
Participants' feedback further offers valuable design insights for future systems aiming to enhance caption writing.
\end{abstract}




\maketitle

\section{Introduction}
Captions for scientific figures, such as charts, bar charts, and pie charts, are crucial to readers.
%It is known that 
Readers comprehend and recall the underlying information significantly better when reading charts with captions, as compared with reading charts alone~\cite{nugent1983deaf,large1995multimedia,bransford1979human,hegarty1993constructing}. Unfortunately, authors do not pay needed attention to crafting figure captions in their papers.
A recent study showed that over half of the figure captions in arXiv papers were rated as not helpful by Ph.D. students~\cite{huang2023summaries}. 
Meanwhile, with advances in deep learning, computational models now have a rich set of capabilities surrounding scientific figures: generating captions of decent quality~\cite{tang2023vistext,masry2023unichart,kantharaj-etal-2022-chart}, analyzing the information in figure images~\cite{kim2021linechart,choi2019visualizing,liu2022matcha}, and evaluating the usefulness of figure captions~\cite{hsu-etal-2023-gpt}.
%However, literature has little to say about how these technologies could be used to aid caption \textit{writers}.
However, literature has little to say about how these technologies could be used to help academics who \textit{write} figure captions for their papers.
The user studies of these technologies were almost all conducted from a \textit{reader} perspective, \ie, having participants read the machine-generated outputs and judge their quality~\cite{yang2023scicap+,huang2023summaries,ye2023mplug}.
The user needs of writers and readers are known to be different~\cite{hinds1987reader,chang2014informal,page1974author,conrad1996investigating}; a piece of machine-generated text might be of lower readability but could serve as a useful draft for writers to work off of~\cite{koehn2009interactive,gaur2016effects,garcia2011machine}.
A few prior works, such as Intentable~\cite{choi2022intentable}, InkSight~\cite{lin2023inksight}, and AutoTitle~\cite{liu2023autotitle}, 
built systems that generate captions considering writers' intentions and allow authors to refine visualization captions and titles, but they did not focus on composing figure captions for scholarly articles.

\begin{figure*}[t]
    \centering
    \includegraphics[width=0.8\textwidth]{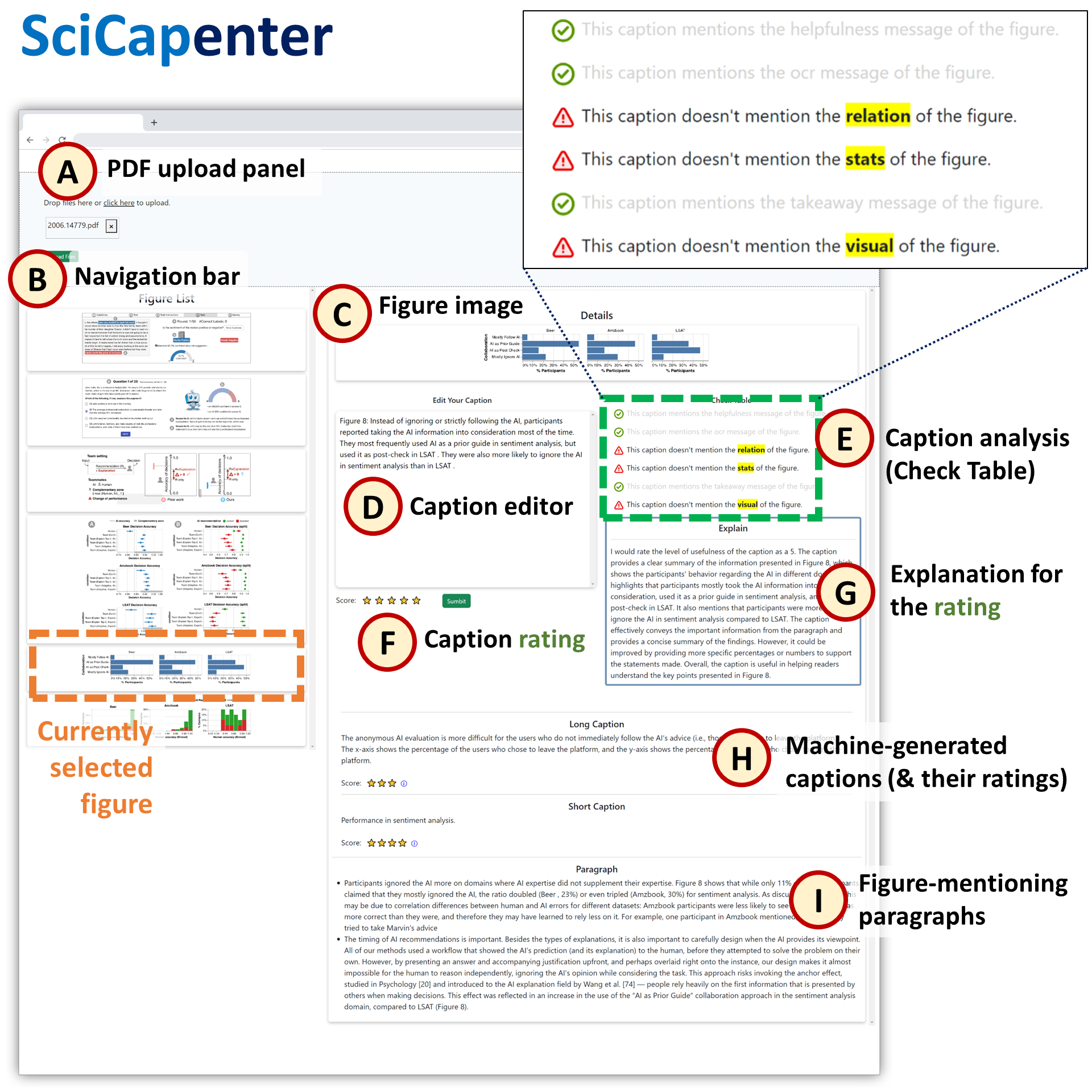}
    \Description[This image depicts the interface of the \system system, which includes several key components for document and caption management.]{}
    \vspace{-.5pc}
    \caption{Overview of \system system interface. 
    \textbf{PDF Upload Panel (A):} A drag-and-drop interface for uploading PDF files.
    \textbf{Navigation Bar (B):} A horizontal bar showing a list of figures extracted from the uploaded document.
    %\textbf{Currently Selected Figure (Indicator):} A visual highlight showing the currently active or selected figure within the interface.
    \textbf{Figure Image (C):} The main area displaying the image of the selected figure.
    \textbf{Caption Editor (D):} A text box for editing the caption of the selected figure.
    \textbf{Caption Rating (F):} A feedback system that allows GPT to rate the quality of the caption, represented by a star rating.
    \textbf{Caption Analysis (Check Table) (E):} Icons indicating the presence or absence of key elements in the caption, such as helpfulness or takeaway message.
    \textbf{Explanation for the Rating (G):} A textual explanation providing insight into why a particular star rating was given to the caption.
    \textbf{Machine-generated Captions \& Their Ratings (H):} This section includes long and short captions generated by AI models, each accompanied by their respective star ratings.
    \textbf{Figure-mentioning Paragraphs (I):} Paragraphs in the document that mention the target figure, providing context or additional information.}
    \vspace{-1pc}
    \label{fig:system-overview}
\end{figure*}

This paper introduces \system, a system that puts together a set of cutting-edge AI technologies surrounding scientific figures to help users write figure captions in scholarly articles (Figure~\ref{fig:system-overview}).
Users first upload the PDF file of their working draft to \system, which will automatically extract all the figures and their current captions.
Users can then browse and pinpoint figures they wish to modify.
Upon selecting a figure, \system displays the figure image, its current caption, paragraphs in the article referencing the exact figure (\eg, ``Figure 3 demonstrates...''), and a series of AI-generated insights. Specifically:

\begin{itemize}
    
    \item 
\system offers two \textbf{machine-generated captions} (Figure~\ref{fig:system-overview}.H): one long, another short, using a cutting-edge AI model. 
We decided to provide captions with two different lengths to address the dual needs of caption writers: longer captions are more informative for readers~\cite{huang2023summaries}, while shorter captions help authors meet space constraints in papers.

\item 
\system provides a \textbf{checklist} (Figure~\ref{fig:system-overview}.E), predicted by an AI model, detailing essential caption features (\eg, does the caption highlight the central message? Does it reference figure text?).

\item
\system \textbf{rates} all captions (Figure~\ref{fig:system-overview}.F), including the author's original and all AI-generated ones, and also shows the \textbf{rationale behind each rating} (Figure~\ref{fig:system-overview}.G). Users can modify captions directly in \system (Figure~\ref{fig:system-overview}.D), submit them for updated ratings, and refine them iteratively.
\end{itemize}

A user study involving 15 STEM Ph.D. students spanning Computer Science to Mechanical Engineering indicated that \system significantly reduced the cognitive load of caption writing, measured by the NASA task load index (NASA TLX).
The majority expressed a desire for similar tools in their future works. 

The contribution of this work is three-fold.
First, our work demonstrates that even if machine-generated text is not superior to human-written text, it could still aid writing.
Second, the user study results highlight the advantages of presenting automated predictions, such as target text's characteristics, when providing writing assistance.
Finally, participants' feedback offers valuable design insights for future systems aiming to enhance caption writing.

\section{Background}

Figures in documents, such as bar charts, line charts, and pie charts, often contain key information the authors desire to convey.
To fully decode the message embedded in figures, reasoning across languages (\eg, document texts, figure captions) and vision (\eg, figure images) is required.
%This poses a significant challenge to the AI community, attracting numerous researchers to develop technologies specifically to address figures.
This complex task presents a great challenge for the AI community, attracting researchers to develop technologies specifically to address figures.
%Many researchers built technologies particularly to handle figures.
%computational models aiming to parse, generate, or evaluate figures and their captions.
%that could advance AI, attracting many 
%A series of technologies were developed to process figures.
%that aim to process figures have recently emerged and, to some extent, matured. 
%general figure tech
These advancements include AI models that 
parse information like trends, axes, or statistics in figure images~\cite{qian2021generating,kantharaj-etal-2022-chart}, 
generate a variety of descriptions based on the image of the figure (\ie, vision-to-language models)~\cite{masry2023unichart,liu2022matcha,lee2023pix2struct} or the data underlying it (\ie, data-to-text models)~\cite{gong-etal-2019-enhanced,obeid-hoque-2020-chart},
answer questions about figure images~\cite{masry2022chartqa,kahou2017figureqa,kantharaj-etal-2022-opencqa}, 
and suggest appropriate figure types based on the data~\cite{siegel2016figureseer,jobin2019docfigure,karishma2023acl}.

Among these efforts, a significant portion was devoted to the study of scientific figures and their captions in scholarly articles.
The \textsc{SciCap} dataset~\cite{hsu-etal-2021-scicap-generating}, introduced in 2021, was the first large-scale collection of real-world figures and captions, containing 2,170,719 figures and captions extracted from 295,028 arXiv papers.
Unlike previous works that used synthetic data~\cite{kahou2017figureqa,kafle2018dvqa,chen2020figure}, \textsc{SciCap} facilitated the development of AI models capable of handling real-world scientific figures. This includes models for generating captions based on figure images~\cite{hsu-etal-2021-scicap-generating,yang2023scicap+,mahinpei2022linecap,li2023scigraphqa,horawalavithana2023scitune,singh2023figcaps} or paper content~\cite{yang2023scicap+,li2023scigraphqa,horawalavithana2023scitune,huang2023summaries}.
There are also models that evaluate the quality of given captions~\cite{hsu-etal-2023-gpt}.
In October 2023, the first \textsc{SciCap} Challenge was held~\cite{scicapchallenge2023}, where six global teams competed on a performance leaderboard, showcasing the advancements in this field. 

From a Human-Computer Interaction (HCI) perspective, these emerging technologies have great potential for aiding caption writing. 
%However, their application in the context of writing has been infrequent.
However, they were rarely studied in a writing context.
Instead, the focus of most user studies on these technologies has been from a reader's standpoint.
This includes methods like showing participants machine-generated figure captions and asking them to assess their quality~\cite{wang2023making,kim2021towards,stokes2022striking,mahinpei2022linecap},
presenting various caption variations for participants to compare or rank~\cite{wang2023making,kim2021towards,stokes2022striking},
or testing how much information readers can remember after reading a figure or a caption~\cite{kim2021towards,stokes2022striking}.
Given that the needs of writers and readers are distinct~\cite{hinds1987reader,chang2014informal,page1974author,conrad1996investigating}, studies focusing on writing scenarios could greatly enhance the application of these technologies in support of caption writing.
A few studies and systems have been developed to aid caption writing.
For instance, Intentable combines automated caption generation with manual input, empowering authors to guide the captioning process based on their insights and intentions derived from the visualization~\cite{choi2022intentable};
InkSight allows users to highlight areas of interest in visualizations through sketching, then automatically generates documentation that users can further revise~\cite{lin2023inksight}; and 
AutoTitle facilitates the interactive creation of titles for visualizations~\cite{liu2023autotitle}.
However, the focus of these systems was not on supporting figure caption composition in scholarly articles.

In this paper, we introduce \system and delve into exploring this area.

\section{\system System}

\system is a comprehensive system that includes a web interface and a caption editing function.
\Cref{fig:system-overview} shows a screenshot of the entire system.
To begin using \system, users start by uploading their PDF file in the PDF upload panel (\Cref{fig:system-overview}.A).
Once the PDF file is processed, the extracted figures are displayed on the navigation bar (\Cref{fig:system-overview}.B).
Users can click on a figure to access the detailed information.
Upon clicking, the selected figure (\Cref{fig:system-overview}.C), along with complete information, appears on the right side of the window.
On the right side of the window, users have access to several features that assist them in composing captions, including
a checking table (\Cref{fig:system-overview}.E), 
GPT evaluation and explanation (\Cref{fig:system-overview}.F and \ref{fig:system-overview}.G),
generated captions - long/short versions (\Cref{fig:system-overview}.H),
and (4) referred paragraphs (\Cref{fig:system-overview}.I).
Users are able to edit the caption in the caption editor (\Cref{fig:system-overview}.D) and resubmit it for re-evaluation.

In the backend, once a user submits a pdf file, \system will store the file in our MongoDB and extract figures, captions, abstracts, and contents using {\tt pdffigure2}~\cite{clark2016pdffigures}.\footnote{Note that in this study, figures classified as ``Table'' were excluded due to the inherent differences between scientific figures and tables. Despite our belief in \system's ability to help users write captions for tables, we opted to omit tables (and figures classified as Tables) in this study.}
% Note that figures labeled with ``Table'' are removed in this study even though we believe \system could also work for tables.
The extracted figure list will then be presented in the navigation bar (\Cref{fig:system-overview}.B).
Regular expressions are then used to identify the corresponding referred paragraphs (\Cref{fig:system-overview}.I) based on the figure index, similar to how \citeauthor{huang2023summaries} extracted figure-mentioning paragraphs~\cite{huang2023summaries}.

% After user submit the pdf file, it will save in the server. Our system further parse the uploaded pdf file using pdffigures2\cite{clark2016pdffigures} to extract the figures, captions, abstract and contents in the file. To reduce the noise of figures, we filter the type of figures that are Table. We then use regular expression to extracted the referred paragraph based on the figure index from the extracted contents. 

The check table (\Cref{fig:system-overview}.E) detects whether the written caption describes or expresses the following six aspects~\cite{hsu-etal-2023-gpt}:

% \review{it's unclear where six aspects were derived from?}
\begin{enumerate}
    \item \textbf{Helpfulness:} Does this caption help you understand the figure?
    \item \textbf{OCR (Optical Character Recognition):} Does this caption mention any words or phrases that appear in the figure? (Examples include the figure title, X or Y axis titles, legends, names of models, methods, subjects, etc.)
    \item \textbf{Relation:} Does this caption describe a relationship among two or more elements or subjects in the figure? (For example, "A is lower than B," "A is higher than B," or "A is the lowest/highest in the figure.")
    \item \textbf{Stats:} Does this caption mention any statistics or numbers from the figure? (For example, "20\% of..." or "The value of .. is 0.33...".)
    \item \textbf{Takeaway:} Does this caption describe the high-level takeaways, conclusions, or insights the figure tries to convey?
    \item \textbf{Visual:} Does this caption mention any visual characteristics of the figure? (Examples include color, shape, direction, size, position, or opacity of any elements in the figure.)
\end{enumerate}
Missed aspects would be highlighted as a warning for users.
The aspect detection model is a SciBert~\cite{beltagy2019scibert} model fine-tuned on 3,159 human annotations, which achieved an F1 score of 0.64~\cite{hsu-etal-2023-gpt}.
The caption rating and its explanation (\Cref{fig:system-overview}.F and \Cref{fig:system-overview}.G) are obtained by calling OpenAI's GPT-3.5-turbo API~\cite{openai2022gpt35}
% \footnote{https://platform.openai.com/docs/models/gpt-3-5 and we use gpt-3.5-turbo.} %\cy{citation} 
with the prompt shown in Appendix~\ref{sec:appendix-prompt} in a zero-shot manner.
% \vspace{.5pc}
% \texttt{
% \begin{quote}
% Given the paragraph and caption below, please rate the level of usefulness
% of the caption from 1 to 6 based on how well the caption could help readers
% understand the important information. 6 is the highest; 1 is the lowest.
% Please also explain your rating. \newline
% Paragraph: ``[paragraph]'' \newline
% Caption: ``[caption]'' \newline
% \end{quote} \vspace{-.5pc}
% } where \texttt{[paragraph]} and \texttt{[caption]} are the placeholder for the referred paragraph and written caption, respectively.
% Note that we choose GPT-3.5-turbo over GPT-4 due to its higher stability and shorter response time.
The machine-generated captions (\Cref{fig:system-overview}.H) are generated with the model proposed by \citet{huang2023summaries}.
Specifically, we obtained their Pegasus$_{P+O}$ and Pegasus$_{P+O+B}$ model, figure captioning models that generate captions by summarising the figure-mentioning paragraphs.
The Pegasus$_{P+O+B}$ model is trained with captions longer than 30 words so it naturally generates longer captions.
In \system, we use the default decoding strategy (beam search with the number of beams = 5) for text generation.

\section{User Study}

\paragraph{Study Overview and Participant Recruitment.}
To assess the effectiveness of \system in assisting users with writing figure captions for academic articles, we recruited 18 participants--3 for a pilot study and 15 for the main study.
Many design details of the main study were informed by the pilot study, which we describe in Appendix~\ref{app:pilot-study}.

For recruitment, we leveraged various channels: personal networks, social media posts, Slack group channels, and the university's mailing list.
These participants were Ph.D. students from the authors' university, majoring in STEM. 
Specifically, 6 majored in Computer Science, 7 majored in Informatics, 3 majored in Chemical Engineering and 2 were Mechanical Engineering majors.
Ideally, we wanted participants currently writing new papers with figures and captions, but such individuals were scarce.
Asking participants to revise and improve poor-quality captions in their published papers also proved challenging, as it required them to admit their papers' weaknesses, complicating recruitment. 
Consequently, we opted to have participants write figure captions for others' published papers in their fields. 
Although different from writing captions for figures in their own paper, this method struck a balance between ease of recruitment and task relevance, making it a practical choice for this preliminary study.

\begin{figure*}[t]
    \centering
    \includegraphics[width=0.95\textwidth]{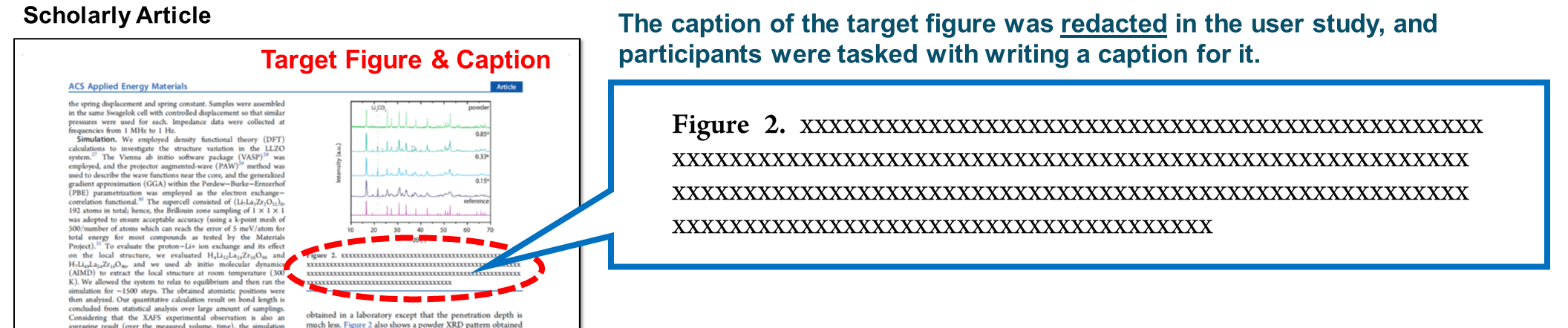}
    \Description[This image provides a visual summary of the preparatory stage of a research study.]{}
    \vspace{-.5pc}
    \caption{Before the study, participants provided ten papers from their research domain, either intended for reading or briefly skimmed but not read in-depth. We processed these papers through \system, choosing six target figures and manually redacting their captions. Participants received the redacted PDFs in the user study and were asked to write captions for the figures.}
    \vspace{-.5pc}
    \label{fig:text-redacted}
\end{figure*}
%(Paper in the image:~\cite{cheng2018garnet})

\paragraph{Pre-Study Preparation.}
We asked each participant to provide ten scientific papers from their research domain
that they intended to read or had briefly skimmed but had not read in-depth.
This approach ensured participants had some contextual understanding of the article
but were not biased by the original captions.
After receiving these articles, we processed them through \system,
selecting six figures with the lowest quality score rated by \system.
%Figures identified as ``Table'' type by \system or those the system failed to parse correctly were excluded.
We then manually redacted the captions from these selected papers using Adobe Acrobat Pro (Figure~\ref{fig:text-redacted}).
All these materials were prepared before the user study.
%In this step, 
We tried our best to select only one figure per article.
However, {\tt pdfparser2} failed to parse papers from some specific domains much more frequently, so this was not always feasible.
In cases where two figures were selected from the same paper for a participant, they received a PDF where both captions were redacted.

\paragraph{Study Procedure and Execution.}
Most sessions were conducted in-lab, with seven participants joining via Zoom.
Each session typically spanned between 1 to 1.5 hours. 
Participants attending in person used one of the author's laptop, while those joining remotely used their own computers.
As \system is a web-based application, it can be accessed anywhere online.
Adobe Acrobat Pro, licensed to all the students in the authors' university, was essential for the study.
Hence, we instructed all the remote participants to install it on their computers beforehand.

The steps of the study were as follows:
We first introduced participants to the study's objective and procedure, securing their consent.
We then offered a brief overview of the \system and its features.
Participants then proceeded through four conditions:
% \cy{Are they writing the same 6 captions for each of the settings below?}

\begin{itemize}

\item 
\textbf{Condition 1 (2 figures): Free writing.} 
Participants were given a PDF with the target caption redacted and asked to draft a caption directly in Adobe Acrobat Pro without any time constraints.

\item
\textbf{Condition 2 (2 figures): Free writing with \system.}
Participants were given a redacted PDF as in Condition 1.
However, this time, they wrote captions directly in \system.
Participants were allowed up to two submissions to retrieve \system's ratings.
The submission limit was set to prevent participants from trying to trick the system just to get higher scores.\footnote{Limiting users to only two attempts for ratings may influence their opinion of the tool and could bias their responses on the NASA task load index. However, our goal was to discourage users from carelessly adjusting their captions without thoughtful consideration. This decision was informed by the writing effort and average caption changes observed in the pilot study.}

\item
\textbf{Condition 3 (1 figure):\footnote{We initially planned for participants to write captions for two figures under all conditions. However, our pilot study indicated that in time-constrained settings, our setup caused excessive effort and rapid fatigue among users. Therefore, we had participants caption only one figure in time-constrained conditions.} Time-constrained writing.}
This condition mirrored Condition 1, but participants were allotted just 8 minutes per figure.
The rationale for this time-constrained setup was to test \system's utility when participants were under pressure for curating captions, such as during last-minute submissions or when captions were given scant attention.

\item
\textbf{Condition 4 (1 figure): Time-constrained writing with \system.} Same as Condition 2, but participants had only 8 minutes to draft their captions.

\end{itemize}

Participants sequentially progressed through the conditions: 1 -> 3 -> 2 -> 4.
This order, derived from our pilot study (Appendix~\ref{app:pilot-study}), helped prevent confusion and fatigue from switching between Adobe Acrobat Pro and \system.\footnote{As participants were new to \system, we understand that this order did not completely remove the learning effect. We believe it makes more sense for participants to first attempt the task without additional support (Condition 1) because writing captions for figures in other people's papers is not a typical writing task.}
After each condition, participants completed the NASA-Task Load Index (NASA-TLX) questionnaire~\cite{hart2006nasa} via a Google Form.
We used a five-point Likert scale (Very Low to Very High) for all scale items, which was used in several prior HCI studies~\cite{lai2022human,mathis2021fast}.
Upon concluding the whole user study, we conducted a brief, open-ended interview to gather feedback and suggestions from each participant. The duration of the post-interview typically varies, but on average, it lasts about ten minutes. We pose four questions to the participants: \textbf{(1) Do you find the system useful, and what aspects do you like or dislike? (2) What is your preferred writing style or approach for figure captions? (3) Do you have any suggestions for our system, and if you could incorporate additional functions, what would they be? (4) Do you have any further feedback?} 
We audio-recorded and transcribed each interview. Then, two authors used open coding to analyze the transcripts.

\section{Findings\label{sec:findings}}

%\ed{We calculate the T-test of total cognitive load for each participants w and w/o system, Free writing condition, P = 0.0459, Time-constrained condition, P = 0.0002}

\paragraph{\system significantly reduced cognitive loads in caption writing for participants, particularly under time constraints.}
%\subsection{Cognitive Load}
In the main study, there were a total of 15 participants.
We computed the NASA-Task Load Index for each condition,
employing a five-point scale ranging from Very Low to Very High and from Perfect to Failure.
The results are shown in \Cref{tab:nasa-task-load}, demonstrating that \system significantly reduced cognitive load for users.
%\cy{Does it pass T-test? ``significantly'' is a very strong word.}
With the assistance of \system, users experienced reduced effort, frustration, and mental demand when composing captions. 
This effect was more pronounced under time constraints, with the overall NASA Task Index (\Cref{tab:nasa-task-load}) dropping from 2.93 to 2.16, compared to a smaller reduction from 2.39 to 2.04 in free writing.
Additionally, the decrease in frustration was the most significant, with the workload falling from 2.33 to 1.40 in free writing, and from 2.73 to 1.67 in time-limited scenarios.

\begin{table*}[t]
\centering
\addtolength{\tabcolsep}{-1mm}
\begin{tabular}{@{}lcccccccccc@{}}
\toprule \normalsize
& \multicolumn{5}{c}{\textbf{Free Writing}} & \multicolumn{5}{c}{\textbf{Time-Constrained Writing}} \\ \cmidrule(lr){2-6} \cmidrule(lr){7-11}
 & \multicolumn{2}{c}{\textbf{User Only}} & \multicolumn{2}{c}{\textbf{With System}} & \multirow{2}{*}{\textbf{p-value}} & \multicolumn{2}{c}{\textbf{User Only}} & \multicolumn{2}{c}{\textbf{With System}} & \multirow{2}{*}{\textbf{p-value}} \\ \cmidrule(lr){2-3} \cmidrule(lr){4-5} \cmidrule(lr){7-8} \cmidrule(lr){9-10}
 & \textbf{Avg.} & \textbf{95\% CI} & \textbf{Avg.} & \textbf{95\% CI} &  & \textbf{Avg.} & \textbf{95\% CI} & \textbf{Avg.} & \textbf{95\% CI} &  \\ \midrule
\textbf{Mental Demand} & 2.93 & {[}2.40, 3.47{]} & 2.53 & {[}1.91, 3.16{]} & 0.233 & 3.80 & {[}3.32, 4.28{]} & 2.53 & {[}1.91, 3.16{]} & <0.001\textsuperscript{***} \\
\textbf{Physical Demand} & 1.87 & {[}1.40, 2.33{]} & 1.93 & {[}1.40, 2.47{]} & 0.774 & 2.20 & {[}1.68, 2.72{]} & 1.87 & {[}1.40, 2.33{]} & 0.019 \\
\textbf{Temporal Demand} & 1.87 & {[}1.36, 2.37{]} & 1.53 & {[}1.18, 1.89{]} & 0.096 & 3.20 & {[}2.53, 3.87{]} & 2.40 & {[}1.74, 3.06{]} & 0.005\textsuperscript{**} \\
\textbf{Performance} & 2.47 & {[}1.88, 3.05{]} & 2.47 & {[}1.75, 3.19{]} & 1.000 & 2.73 & {[}2.12, 3.34{]} & 2.13 & {[}1.48, 2.79{]} & 0.095 \\
\textbf{Effort} & 2.87 & {[}2.36, 3.37{]} & 2.40 & {[}1.90, 2.90{]} & 0.150 & 2.93 & {[}2.40, 3.47{]} & 2.33 & {[}1.84, 2.83{]} & 0.057 \\
\textbf{Frustration} & 2.33 & {[}1.62, 3.05{]} & 1.40 & {[}1.12, 1.68{]} & 0.008\textsuperscript{**} & 2.73 & {[}2.06, 3.41{]} & 1.67 & {[}1.21, 2.12{]} & 0.001\textsuperscript{**} \\ \hline
\textbf{Overall} & 2.39 & {[}2.05, 2.73{]} & 2.04 & {[}1.71, 2.38{]} & 0.046\textsuperscript{*} & 2.93 & {[}2.52, 3.34{]} & 2.16 & {[}1.74, 2.57{]} & <0.001\textsuperscript{***} \\
\bottomrule
\end{tabular}
\caption{We leverage the NASA Task Load Index method to assess workload on five-point scales, under conditions of free writing and time-constrained writing, both with and without \system. A lower score indicates a lower perceived workload. The p-values are obtained by comparing the user-only and with-system settings (paired t-test, two tailed, N=15). The 95\% confidence intervals are estimated using t-distribution. With \system, users had less workload in both free writing and time-constrained writing scenarios, with the most notable improvements seen in the Mental Demand and Frustration.}
\vspace{-1.5pc}
%kenneth{TODO CY and Ed: Use this new table. (0) MAKE SURE THE NUMBER ARE CORRECT!, (1) Replace the numbers, (2) add standard deviation or other ways of expression error interval, and (3) Enrich captions to say the take away message.}\ed{done except (2)}
\label{tab:nasa-task-load}
\end{table*}
\addtolength{\tabcolsep}{1mm}

\begin{figure*}[t]
\centering
\includegraphics[width=0.4\linewidth]{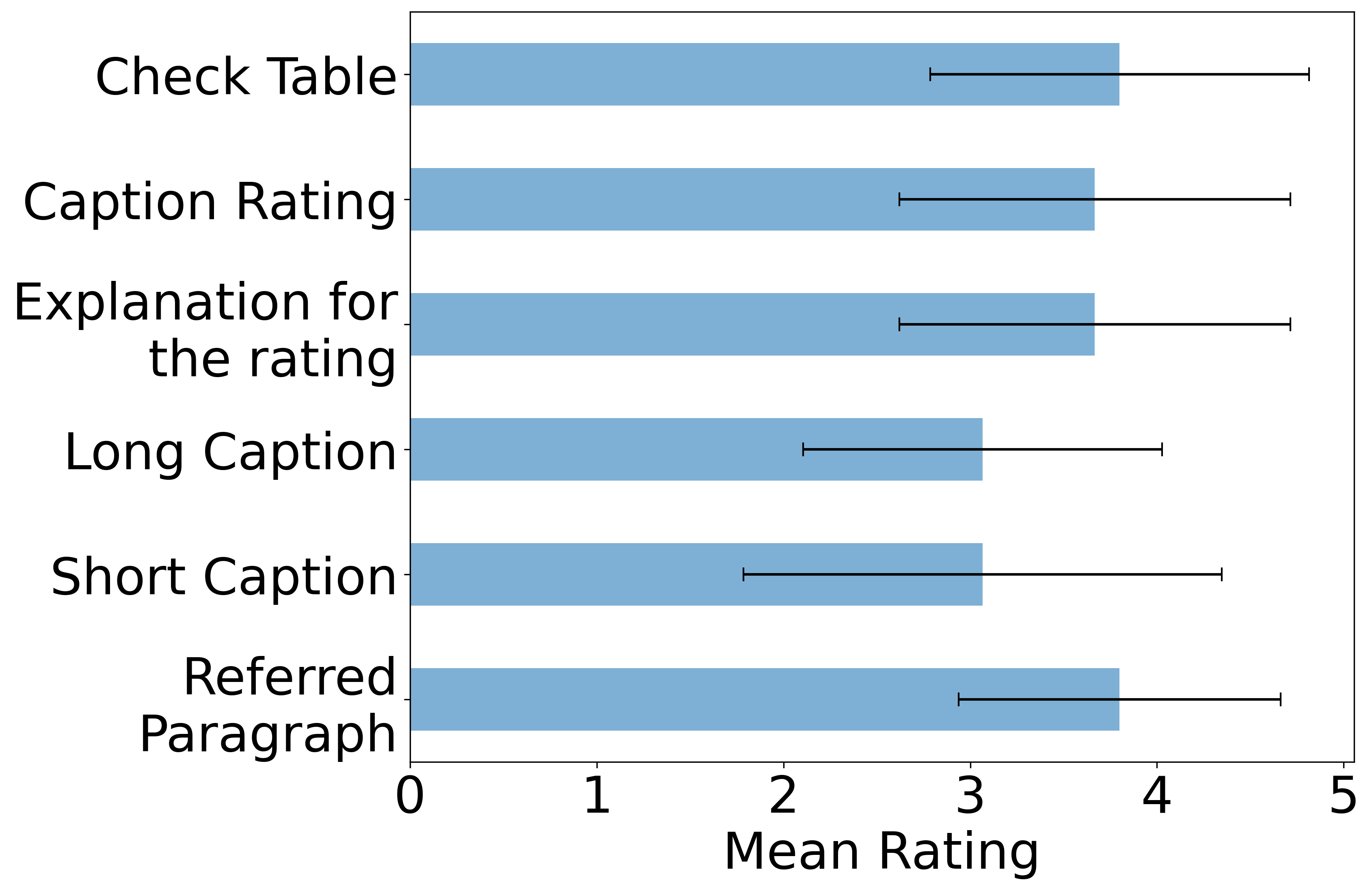}
\includegraphics[width=0.54\linewidth]{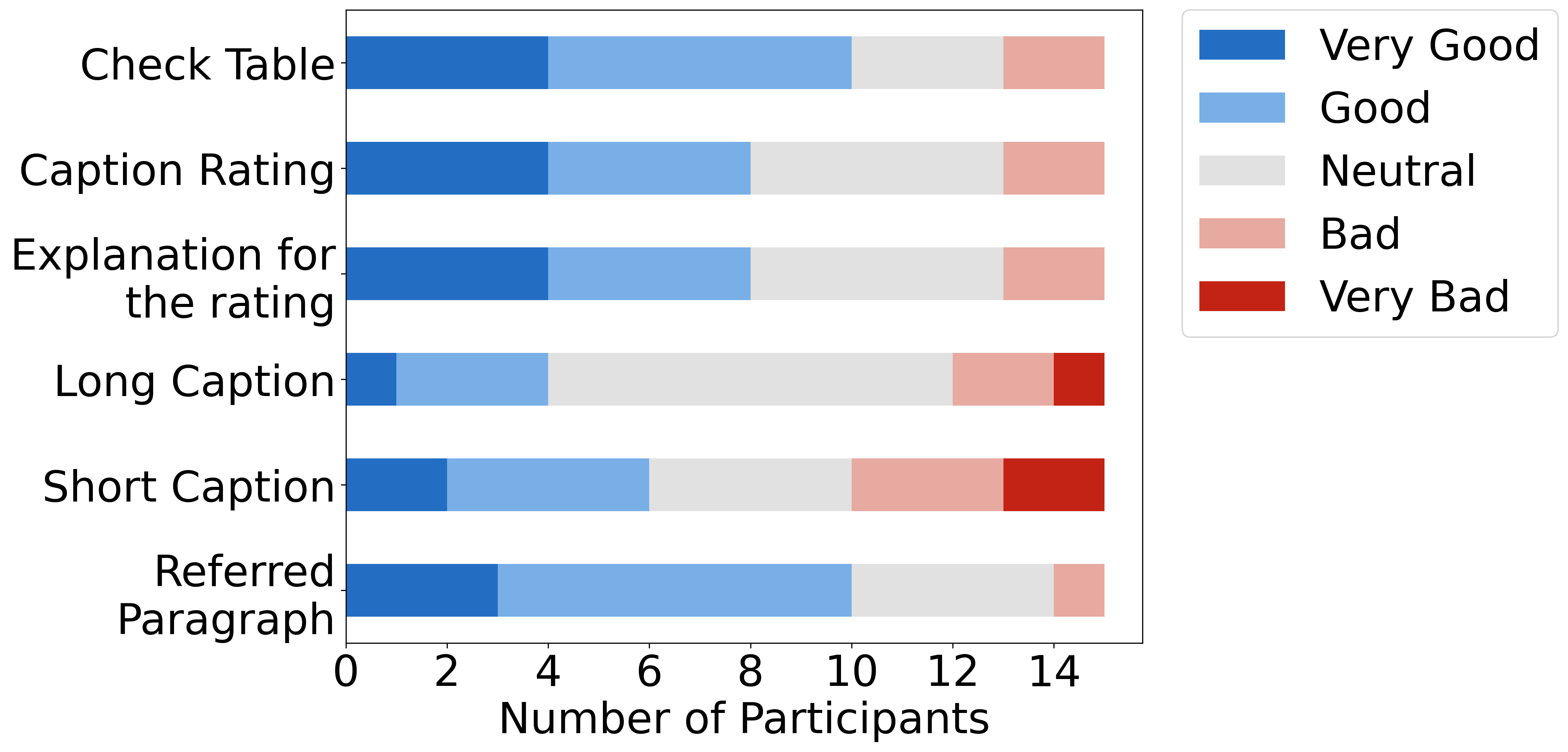}
\Description[The image compares six \system features, showing higher ratings for check tables and paragraphs, and lower for captions; the short caption's ratings are particularly varied.]{}
\vspace{-.5pc}
\caption{Comparison between six different elements provided by \system. Left figure shows the mean rating and standard deviation of a five-point scale for different elements. The check table and referred paragraph were rated highest, while short and long caption had the lowest score, indicating they were the least favored elements according to the participants. 
    Right figure shows a breakdown of the five-point scale with different colors representing each rating. The short caption exhibit a more varied distribution of opinions.}
    \vspace{-.5pc}
    \label{fig:enter-label}
\end{figure*}

\paragraph{Machine-generated ratings and labels were deemed more useful than machine-generated captions.}
%\subsection{Ratings Toward \system}
We asked participants to rate the overall \system in terms of the following three questions on a five-point Likert scale:
{\em (i)} How difficult is it to use the system? (Rating scale: Very Easy to Very Difficult),
{\em (ii)} How useful is the system in assisting with caption writing? (Overall) (Rating scale: Very Bad to Very Good), and
{\em (iii)} How fast is the system response? (Rating scale: Very Slow to Very Fast).
Six more questions about the satisfaction level (a five-point Likert scale from Very Bad to Very Good) toward the six components (\ie, check table, caption rating, explanation for the rating, long caption, short caption, and referred paragraph) were also asked.
Figure~\ref{fig:enter-label} shows the results.

Overall, participants expressed high satisfaction with our system's usability (average rating = 3.80, SD = 0.86, n = 15) and its ease of use (average rating = 1.80, SD = 0.68, n = 15).
The response time got an average rating of 3.40 and was not considered too slow.
% The detailed results are presented in \Cref{tab:system-like-scale}. 
We also discovered that, within \system, participants found machine-generated ratings and predictions more useful than machine-generated captions. 
Even extracted paragraphs that mentioned the figure were deemed more helpful.
Both long and short captions had the lowest average preference rating.
Analyzing the ratings further, long and short captions received the highest number of ``Very Bad'' ratings. %\kenneth{Is this true???}
The post-study interview of participants suggested that the system-generated captions often fall short of expectations, with some experiencing inaccuracies or ``hallucinations'' (P8, P10, P11, P15), especially short captions:
``I think the generation wasn't 100\% precise... sometimes it captures some weird stuff. (P8)'',
``the long and short caption... it wasn't that good. (P10)'',
``So, for the generated caption, why is not useful? Because it just have [sic] some hallucination, or they just write what you don't want. ...I don't think they are explaining to [sic] figure well. (P15)''

 %\textit{P8 -  I think the generation wasn't 100\% precise. Of course not, but like, sometimes it captures some weird stuff. But it's also nice that some keywords could also be used, like just describe when you're writing}

    %\textit{P10 - the long and short caption, I think maybe because of our reason or something, it wasn't that good}

    %\textit{P11 - for the long captions, I don't think it's useful because for my personal preference I don't like the long captions, but for the short captions, sometimes it doesn't meet my requirement}

    %\textit{P15 - So for the generated caption, why is not useful? Because it just have some hallucination, or they just write what you don't want? Sometimes it had some hallucinations and sometimes they were not useful. I don't think they are explaining to figure well.}

%\{Users have diverse preferences towards different features.}
\paragraph{Users show varied preferences for the features.}
%Participants' interviews reveal varied preferences for each system feature.
Most participants (P1, P4, P7, P10, P11, P14) preferred short captions for conciseness and simplicity, while a few (P3, P13) chose long captions for their detail, particularly with complex or important figures.
The effectiveness of the check table also differed among participants. 
Some valued its directive nature for caption content (P3, P4, P5, P7, P15):
``the good side about it (\textsc{SciCapenter}) is it has a check table, so it kind of reminds me which element I should put or which criteria I should be checking before I write the caption or while I'm writing the caption. (P3)'',
``the check table is really helpful because I can know what information is missing in my caption. (P4)''
However, some found it confusing or irrelevant (P9, P11, P12):
``I don't know how to use this check table to improve my caption because it sometimes says you need to add stats for the caption, but for the figure, actually I'm not able to add some statics for these figures. (P9)'',
``I don't like that (check table). I don't check every part of it (the caption) because I think in some cases you don't need to cover all the things. (P12)''

Meanwhile, although paragraphs and explanations were typically helpful in understanding and describing figures (P2, P3, P6, P15), there were instances where explanations failed to reflect the participants' recent edits (P4, P8).

%\paragraph{Concrete suggestions for writing actions was desired.}
\paragraph{Users desire actionable suggestions, templates, or examples for implementation.}
The post-study interviews show a frequent request for the system to offer more concrete action suggestions for easy implementation:
``You want not only the check table, but also give you a suggestion on what to write for the aspects. ...
%Just like the explanation part, maybe you can give me some suggestions about how to improve my caption. Because 
sometimes I saw the check table, but I don't know how to use this check table to improve my caption. (P9)'',
``..., like maybe for each point (in the check table), you can list a sentence or two sentences that can be added to the caption directly,
%or users can modify a referenced recommendation
... So users don't need to come up by themselves. (P10)''.
%``Instead of long, short, and paragraph, I wanted one more that says, okay, the input that you gave, how can I make that itself better? (P13)''
%``You can give me hints-- like bullet points-- so that I can, when I do my paper, I can just follow the guidelines of the systems to finish all the writing... (P11)''
Some suggested \system to provide example captions for similar figures:
``... you can generate a typical caption from other papers that have already been published. (P7)'',
``I would be curious about how other people do for similar figures. ..., I would like to see what people mention in similar domains of papers for similar formats of figures. (P8)''
Some desired a template that they could simply fill in information:
``..., you have standard (templates), and you just need to fill in the details. (P7)'',
``..., you want some, like standard structure. Just a structure, not the exact paragraphs that you need to mention, ... (P11)''.

\section{Discussion}

\paragraph{Design Recommendations.}
Our research indicates that while machine-generated texts may not always match the quality of human-written content, they might still offer assistance in writing tasks.
To address the potential shortcomings of machine-generated texts, \system incorporated an abundance of contextual information, such as caption traits, caption ratings, explanations for these ratings, and references to relevant figures, to aid users.
Based on our study results and user interviews, we offer two design recommendations for writing assistants targeting figure captions:

\begin{itemize}

\item
\textbf{Present contextual or analytical information relevant to the writing task.}
While machine-generated captions were generally deemed useful, our results suggested that the benefits of presenting contextual (\eg, figure-mentioning paragraphs) or analytical (\eg, check tables, ratings) information relevant to the writing task might be more prevalent (Figure~\ref{fig:enter-label}).
%Writing assistants aiming to support caption writing are recommended to present this information.
%As figure captions are highly contextual, 
%If time or resources are limited, consider prioritizing the development of information extractors and analyzers over text generators, as they may be more useful to users.

\item
\textbf{Provide concrete, actionable suggestions for writing.}
\system analyzed the current situation without offering specific guidance on enhancing a caption or providing fillable caption templates, leading to many participant complaints (Section~\ref{sec:findings}). 
Future developers of writing assistants for caption writing should aim to provide more targeted, actionable suggestions. %, such as precise revision tips or fillable templates. 
Continued research and development are crucial for this.

\end{itemize}

\begin{figure*}[t]
    \centering
    \includegraphics[width=0.98\linewidth]{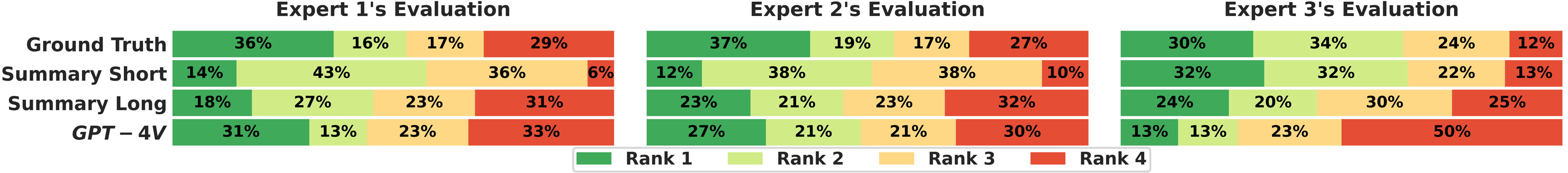}
    \Description[The image shows expert ratings on caption quality for four types: Ground Truth, Summary Short, Long, and GPT-4V, from best to worst. Ground Truth is often top-rated, but experts differ, especially on GPT-4V's performance.]{}
    \caption{Comparative evaluation of caption quality by three experts, where each caption type—Ground Truth, Summary Short, Summary Long, and GPT-4V—is rated on a scale from rank 1 (highest) to rank 4 (lowest). Both Expert 1 and Expert 2 rated Ground Truth caption as rank 1 most frequently, while Expert 3 had a preference for Summary Short. Notably, Expert 3 rated GPT-4V the lowest, rarely giving it a rank 1, whereas both Expert 1 and Expert 2 often considered GPT-4V as their second choice for rank 1. The variations in evaluations reflect differing perspectives on caption quality and suggest that while Ground Truth captions are generally preferred, there's a significant disparity in how each expert rates the machine-generated captions.}
    \label{fig:upwork-result}
\end{figure*}

\paragraph{Comparison of Caption Quality.}
The study was conducted before we had access to GPT-4V~\cite{2023GPT4VisionSC}, one of the strongest Large Vision-Language Models (LVLM), raising an interesting question:
\textbf{Could GPT-4V produce significantly better captions than the models used in \system, potentially altering our study's conclusions?}
To explore this, we recruited three professional academic paper editors to assess the quality of captions generated by various models.

From the SciCap Challenge dataset~\cite{scicapchallenge2023}, we randomly selected 200 figure-caption pairs. 
For each figure, we generated two captions using models from \system (Summary Short and Summary Long), one caption using GPT-4V by prompting it with the figure image and figure-mentioning paragraphs, and we also included the original author-written caption (Ground Truth). 
To ensure a fair comparison in our study, GPT-4V was prompted to generate captions no longer than the author-written captions, acknowledging that readers often favor longer captions. 
This length constraint aimed to minimize the influence of caption length on quality assessment.

We recruited three professional editors (Expert 1, 2, and 3) through UpWork,\footnote{UpWork: https://www.upwork.com/} all specializing in technical academic articles and native American English speakers.
Their backgrounds include one with over ten years of editing experience and a Ph.D. in Comparative Literature, and two from the STEM fields---one in Theoretical Astrophysics and another in Neuroscience---with extensive experience in editing, proofreading, and publishing academic papers. 
They ranked the captions based on their effectiveness in clarifying the figure's intended message. 
The results are shown in \Cref{fig:upwork-result}.
%As a result, 
Under the caption length constraints, GPT-4V did not outperform \system's models drastically.
Although Experts 1 and 2 preferred GPT-4V's generation over that of \system's models, Expert 3 did not favor it.

\paragraph{Limitations.}
Our work demonstrates that AI technologies can support caption writing, yet we acknowledge the following limitations:

\begin{enumerate}
    \item Our study involved participants writing figure captions for other people's published papers rather than crafting captions for their own work. 
This scenario diverged from typical use cases of figure caption writing support systems. 
%as authors possess in-depth knowledge of their paper's context and the relevant scientific background. 
The study was designed this way to accommodate the difficulty in recruiting individuals who are actively writing new scholarly articles. 

\item Despite efforts to mitigate learning effects by altering the condition order based on the pilot study (Appendix~\ref{app:pilot-study}), our study procedure could still introduce potential biases. 
For instance, participants had to learn the uncommon task of writing captions for other people's papers in the first condition they encountered.
%Our study design balanced feasibility with realism in simulating realistic writing scenarios for a controlled user study, acknowledging inherent biases.
%This study design also could not eliminate other biases.
%For example, 
Furthermore, all participants first wrote the captions without \system and then used \system, which may have amplified the system's benefits.

\item 
Our study only used NASA-TLX.
Future research can extend the measures to not only NASA-TLX but to other measures, such as a sense of authority or confidence level, to understand the benefit of \system in multiple aspects.

\item As a preliminary study and early prototype, the lack of consideration for specific items from the Check Table in both generated captions and explanations for ratings suggests a need for better integration with the tool's features.

\item 
Our study %, focusing on the user's cognitive load, 
did not assess the quality of the resulting captions, leaving unanswered whether \system improved caption quality. 
This was due to the high cost and need for domain experts to evaluate caption usefulness.
Further studies are needed to determine writing assistants' impact on caption quality. 

\end{enumerate}

\section{Conclusion and Future Work}
In this paper, we present \system, a system designed to assist authors in crafting captions for scientific figures within scholarly articles.
%The intuition behind \system is that while machine-generated captions may not be flawless, they can still be used to aid writing when presented with rich contextual information.
%significantly enhanced by integrating ample contextual information. 
Given a scholarly article, \system identifies and extracts figures, isolates paragraphs that mention these figures, generates captions based on these paragraphs, and then rates the quality of these captions.
Our user study showed that \system reduced users' cognitive load while composing figure captions.
Considering this is an initial effort in aiding caption creation, our future goal is to develop a more accessible version of \system, potentially in the form of a web browser plugin.
We aim to conduct a broader deployment study to understand how researchers might integrate this tool into their academic writing.
%Another avenue we are keen on exploring is extracting visual information from figure images to guide users in producing even more refined captions, given that the current iteration of \system relies solely on text from scholarly articles.

\bibliographystyle{Style/ACM-Reference-Format}
\bibliography{Bibtex/sample-base,Bibtex/software,Bibtex/main}

% \clearpage

%\section*{Appendix}
\appendix
%\label{sec:appendix}

%\vspace{.5pc}

\section{Prompts Used}
\label{sec:appendix-prompt}

The following is the prompt used by \system to generate
the caption rating and its explanation (\Cref{fig:system-overview}.F and \Cref{fig:system-overview}.G): 

\texttt{
\begin{quote}
Given the paragraph and caption below, please rate the level of usefulness
of the caption from 1 to 6 based on how well the caption could help readers
understand the important information. 6 is the highest; 1 is the lowest.
Please also explain your rating. \newline
Paragraph: ``[paragraph]'' \newline
Caption: ``[caption]'' \newline
\end{quote} \vspace{-.5pc}
} where \texttt{[paragraph]} and \texttt{[caption]} are the placeholder for the referred paragraph and written caption, respectively.
Note that we choose GPT-3.5-turbo over GPT-4 due to its higher stability and shorter response time.

\section{Pilot Study and Changes Made}
\label{app:pilot-study}

The first three study trials were treated as the pilot study.
Two of the participants majored in Informatics, and one in Computer Science.
The pilot study yielded several valuable insights that informed adjustments made in our main study:

\begin{itemize}
    \item 
%\begin{enumerate
In the pilot study, the task sequence was ordered 2 -> 4 -> 1 -> 3, with the \system conditions proceeding first. 
We observed that after participants viewed the six aspects reported in the check table (\Cref{fig:system-overview}.E), their writings were influenced, for example, to include takeaway messages and figures' visual features. 
To minimize this influence, we changed the task order by asking participants to proceed with the conditions without \system first in the formal study.
We also included extra instructions clarifying that the six aspects are suggestions, not requirements.
%To minimize the influence, we decided to change the task order by asking participants to proceed with the conditions without \system first.

\item 
In the pilot study, the rating's explanation (\Cref{fig:system-overview}.G) was only visible when users hovered over a small button.
We found that participants frequently ignored this explanation.
As a result,
we modified the interface to display the explanation directly.
%explicitly under the reference table.

\item
The missing aspects in the check table (\Cref{fig:system-overview}.E) were originally tagged with the icon ``X''.
Users expressed concerns about such a strong rejection symbol. 
We, therefore, replaced it with a triangle alert symbol.

\item 
The time constraint for Condition 3 and Condition 4 in the pilot study was set to five minutes. 
%However, 
We found that participants were quite stressed and could barely finish the caption.
So, we extended the time limit to 8 minutes.

\item
GPT-4 was used for automatic caption evaluation (\Cref{fig:system-overview}.F and \Cref{fig:system-overview}.G) in the pilot study.
However, the API's response time was quite high.
Thus, we changed to GPT-3.5-turbo in the main study.

\end{itemize}

\end{document}